\def\preprint{0}                
\def\preprint{1}                
\def\comment#1{}
\if\preprint1
        \documentclass[usenatbib]{mn2e}
        \usepackage{times}
        \usepackage{graphicx}
        \usepackage{calc}

\else
        \documentstyle[astrop-bib,referee,times]{mn}
        \newcommand{\includegraphics}[1]{}
\fi

\def\oversim#1#2{\lower0.5pt\vbox{\baselineskip0pt \lineskip-0.5pt
     \ialign{$\mathsurround0pt #1\hfil##\hfil$\crcr#2\crcr\sim\crcr}}}

\title[The trigger of the AGB superwind]{The trigger of the  AGB superwind:
 the importance of carbon}
\author[E. Lagadec \& A.A. Zijlstra]{Eric Lagadec \thanks{E-mail:
eric.lagadec@manchester.ac.uk}\&
               Albert~A.~Zijlstra
\\
Jodrell Bank Centre for Astrophysics, Alan Turing Building, The University of Manchester, 
           Manchester M13 9PL, UK\\}
\pubyear{2008}

\begin{document}

\date{Accepted . Received}

\pagerange{\pageref{firstpage}--\pageref{lastpage}} \pubyear{2008}

\maketitle

\label{firstpage}

\begin{abstract}
The driving mechanism of the AGB superwind has become controversial in
recent years. The efficacy of dust-driven mass loss has been
queried. Spitzer observation of AGB stars in Local Group Galaxies 
show the surprising result that at low metallicity,  AGB mass loss occurs
at low luminosity, possibly lower than in the Galaxy, but only for
carbon-rich stars.  Oxygen-rich stars in the Galaxy and in lower
metallicity galaxies have similar mass-loss rates only at high
luminosities.  To explain this dichotomy, we propose that the
superwind has a dual trigger. The superwind starts either when
sufficient excess carbon builds up for efficient formation of
carbonaceous dust (which we propose occurs when $X_{\rm CO} = \rm (C -
O)/ O_\odot = 0.1$), or when the luminosity reaches a value sufficient
for a silicate-dust-driven wind (proposed at $L = 10^4 Z^{-4/3}\,\rm
L_\odot)$. We show that this dual trigger fits the current
observational constraints: the luminosity at which the superwind
begins, and the predominance of carbon superwind star at low
metallicity.  We use stellar evolution models to check the consistency
of our explanations and present detailed predictions of the
luminosities at which the superwind is triggered for different
metallicities and initial stellar masses.
  
\end{abstract}



\section{Introduction}
The late phases of stellar evolution are crucial  to
galactic evolution. Stars in this phase eject much of their gas into
space, enriched by the products of nuclear burning and newly formed
dust. Of all the stars which have died in the Universe so far, over
90\% are low and intermediate-mass stars ($\rm M\sim 1$--$8\,\rm
M_\odot$: LIMS).  In consequence, LIMS are important contributors to the enrichment of the Interstellar Medium
(ISM) by  sources internal to the galaxy.
 
LIMS eject between 20 and 80 per cent of their mass during the
so-called {\it superwind} phase which terminates the high luminosity
asymptotic giant branch (AGB). This near-catastrophic but
non-explosive wind removes effectively the entire hydrogen-rich
envelope. During the superwind, dust forms in the ejecta, and the star
tends to become self-obscured.  After this final burst of activity,
the star remains as a hot, compact white dwarf. The expanding
circumstellar ejecta become transparent, and briefly form an ionised
planetary nebula, before dispersing into the interstellar medium.

At the low effective temperature of AGB stars, the ejecta are largely
molecular.  The first stable and abundant molecule to form is CO: this
locks up the least abundant of the C and O atoms. Other molecules and
dust form from the remaining atoms. Dredge-up of primary carbon
(produced by triple-$\alpha$ burning) causes the C/O ratio to increase
over time.  The chemistry changes dramatically when the C/O ratio
passes unity, and this affects both the gas-phase species (Millar et
al. 2000, Willacy et al. 1997) and the dust composition (Treffers et
al. 1974).  In carbon stars, the C/O ratio (by number) exceeds unity,
and carbon-rich molecules (e.g. C$_2$H$_2$, HCN) and carbonaceous dust
result, whilst in oxygen stars (C/O$<1$), metal oxides and silicate
dust form. This dichotomy is not complete: in the inner shell of AGB stars 
shocks can lead to the formation of some molecules of the 'wrong' type
(Decin et al. 2008).

The mass-loss mechanism of AGB stars has become controversial.  A
two-step process is likely involved. First, shocks due to pulsations
from the star extend the atmosphere (Jones et al., 1981) leading to
dust formation.  Second, radiation pressure on the dust drives an
outflow; the gas is carried along through friction with the dust
particles. Pulsations can drive mass-loss rates up to about
$10^{-6}M_{\odot}$yr$^{-1}$; the much higher rates observed are
explained by dust-driven winds (Bowen \&\ Wilson 1991).  This simple
distinction between weak pulsation winds and dusty superwinds (Winters
et al. 2000) has recently been disputed. Woitke (2006) find in their
models that the dust opacity of silicates is insufficient to drive a
wind, while for carbon dust the winds become very chaotic (Woitke \&
Niccolini 2005).  For low-metallicity stars, dust-driven winds may
 fail due to the lack of dust (Zijlstra 2004): theoretical work by
Bowen \& Willson (1991) predicts that for metallicities below
[Fe/H]$=-1$ the wind becomes pulsation-driven.  Mattson et al. (2008)
argue that pulsation can drive a strong wind at low metallicity, set by
the pulsation energy of the star. Recent theoretical models produce
similar mass-loss rates for AGB stars with SMC, LMC and solar
metallicities (Wachter et al. 2008).

The effect of dust should be most noticeable when comparing stars at
different metallicity. This has become possible with the {\it Spitzer
Space Telescope} (Werner et al. 2004), which provided the sensitivity
necessary to obtain mid-infrared spectra and derive mass-loss rates
for the whole range of AGB masses and luminosities in the Magellanic
Clouds and a few other Local Group Galaxies. These galaxies trace in
general metal-poor stars with a range of metallicity and ages, and the
known distances means that absolute mass-loss rates and luminosities
can be established. The expectation for reduced mass loss efficiency
at low Z has not been confirmed.  Instead, the {\it Spitzer}
surveys have shown that high mass loss rates with dusty winds can be
reached even at very low $Z$, as seen in the Fornax dSph galaxy at
[Fe/H]$\sim -1$ (Lagadec et al. 2008).

One unexpected result of the Spitzer surveys is the almost complete
dominance of carbon stars among the mass-losing stars in these low
metallicity galaxies.  The estimated mass-loss rates for these carbon
stars are very similar to the ones observed in Galactic AGB
stars, while the few oxygen-rich AGB stars observed have mass-loss
rates one or two orders of magnitude lower than their Galactic
counterparts (Sloan et al. 2008).  All this seems to indicate that the
abundance of carbon and oxygen is an important parameter for the
mass-loss from AGB stars at low metallicity.  We will discuss in
the present letter at what point during the stellar evolution the
superwind begins, and how it is triggered.

\section{The AGB stars sample, luminosity and mass-loss rates}
\subsection{Sample selection}

\begin{figure}
\includegraphics[width=9cm]{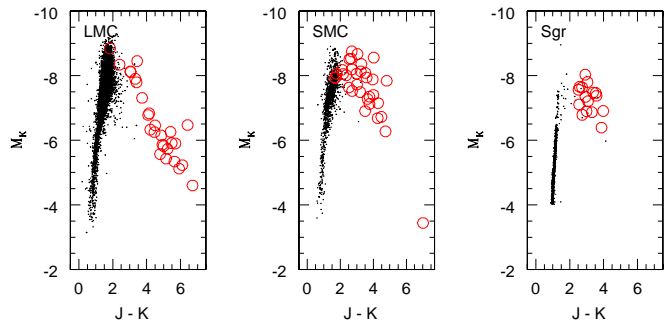}
\caption{\label{jhk} J$-$K vs M$_{\rm K}$ for the LMC and SMC samples.
The dots indicate 2MASS field stars in each galaxy. For the
LMC and SMC, all field carbon stars are shown. For Sgr dSph, a cleaned field
star population (oxygen rich or carbon rich) towards the central square degree 
of the galaxy are shown: see text. }
\end{figure}

The stars from our sample were selected as reddened AGB stars in a
2MASS M$_k$vs J-K diagram. Fig.\ref{jhk} shows the colour--magnitude
diagrams for the Large Magellanic Cloud, Small Magellanic Cloud and
Sagittarius dwarf spheroidal galaxy (hereinafter LMC, SMC and Sgr
dSph).  For the LMC and SMC, the dots indicate optically
identified carbon stars. For Sgr dSph, no such catalogue is
known. Instead we selected all stars within a 1-degree region towards
the galaxy, and removed any star with a close match in (K, J$-$K) with
a neighbouring field. We finally used a colour criterion to remove the
remaining sequences corresponding to foreground stars, to the blue of
the Sgr sequence. The majority of the remaining stars will be
oxygen-rich RGB stars: the AGB is visible as the less populated upper
range of the sequence.  In all three plots, open circles indicate
stars selected for Spitzer observations. For Sgr dSph, these are
selected from a larger area than the plotted unreddened stars.

All three systems show a range of metallicities. We take as
representative values [Fe/H]$\, = -0.3$ for the LMC, $-0.7$ for the
SMC and $-0.55$ for Sgr dSph. The distance moduli are taken as 18.54,
18.93 and 17.02, respectively.

Several different Spitzer programs have obtained mid-infrared
spectroscopy of AGB stars in the Local Group. The description below
refers mainly to works published in Zijlstra et al. (2006); Matsuura
et al.  (2006); Groenewegen et al. (2007); Lagadec et al. (2007) and
Sloan et al. (2006; 2008) for carbon rich stars in the the Magellanic
Clouds (27 stars in the LMC and 33 in the SMC) and van Loon et
al. (1999) for the oxygen-rich stars in the LMC (26 objects).
Matsuura et al. (2007) for 3 AGB stars Fornax and Lagadec et
al. (2008) for 6 others in the Sagittarius Dwarf spheroidal galaxy.

\subsection{Luminosities}

The important onset of the superwind is best traced by the luminosity
distribution of the dusty stars, compared to those of unreddened
AGB stars.

\begin{figure}
\includegraphics[width=9cm]{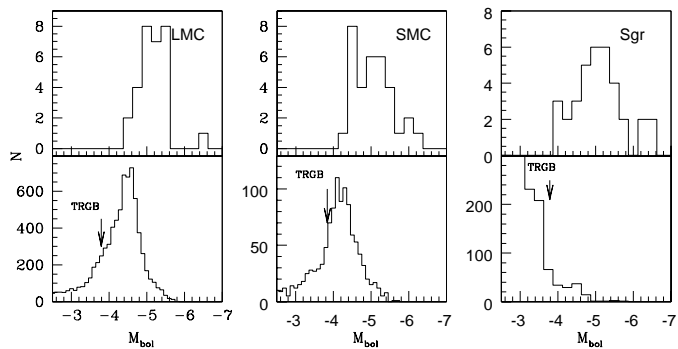}
\caption{\label{mbol} Bolometric magnitudes for  the observed stars (top
panels) and for field  stars. For the LMC and SMC, only field carbon stars
are included. For Sgr dSph, the field population is for
the red colour sequence only, which includes oxygen-rich stars. }
\end{figure}

Fig. \ref{mbol} shows the distribution of the AGB carbon star populations in
the LMC and SMC (bottom panels; left and middle). These were taken from
spectroscopically identified catalogues of carbon stars. The bolometric
magnitudes were derived using the $J-K$, $K$ magnitudes: the bolometric
correction for carbon stars was taken from Whitelock et al. (2006). It is
important to note the uncertainties inherent in such bolometric
corrections. Most of the near-infrared spectral region where the stellar
energy distribution peaks is hidden by our atmosphere. In addition deep
molecular bands are exhibited by the stars. Bolometric corrections derived
from oxygen-rich stars would in general overestimate the luminosity of carbon
stars, as the former corrects for deep water bands which coincide with opaque
spectral regions in our atmosphere.

The histograms of the absolute bolometric magnitudes distributions
for the dusty stars observed with Spitzer, are shown in the upper
panels of Fig. \ref{mbol}. For these, the luminosity was obtained by
integrating over the Spitzer spectra and broadband colours.

We also show some corresponding data for Sgr dSph.  The panel shows
stars selected from the sample of Fig. \ref{jhk}, using a colour
criterion to select the reddest branch seen in the figure. This will
be dominated by oxygen-rich stars.  The observed distribution clearly
shows the dominant inclusion of RGB stars.  We also do not know how
many of the AGB stars are carbon-rich. A comprehensive survey for
carbon stars in Sgr dSph is clearly warranted.

Table \ref{cstars} gives some representative parameters for the carbon
stars in each galaxy, derived from the distributions in
Fig. \ref{mbol}.  The second column lists the magnitude where the
luminosity function of the unreddened carbon stars becomes flat: this
identifies the point where the large majority of AGB stars has become
carbon-rich. The third column indicates the life time of the carbon
star phase: this is derived from the width of the flat part of the
luminosity function, converted to a time scale using the relation
$dM_{\rm bol}/dt = -8.25 \times 10^{-7}
\,\rm mag\, yr^{-1}$ (Wood 1990).  The last column indicates the
bolometric magnitude at which the onset of the superwind occurs: this
is taken as the point where the luminosity distribution of the
mass-losing stars reaches its maximum.  These numbers should be used
with caution: especially the last column is affected by low-number
statistics.

For the LMC, the table and figure show that the mass loss starts
shortly after the star becomes carbon rich. For the SMC, there is a
longer delay. Also, for the LMC, most stars become carbon-rich late on
the AGB. For the SMC, the shift in chemical type coincides with the
tip of the RGB, which is approximately the same as the onset
luminosity for thermal pulses.  Thus,  the SMC stars become
carbon-rich very soon after the thermal pulses begin.

The last column, surprisingly, seems to indicate that the mass loss
starts at a lower luminosity, i.e. earlier on the AGB, for the SMC
than it does for the LMC and the Galaxy ($M_{\rm bol}\approx
-5.5$). One should note that the samples were not selected in quite
the same way, and so any conclusion would require more work.

\subsection{Mass-loss rates}
\begin{figure*}
\includegraphics[width=16cm]{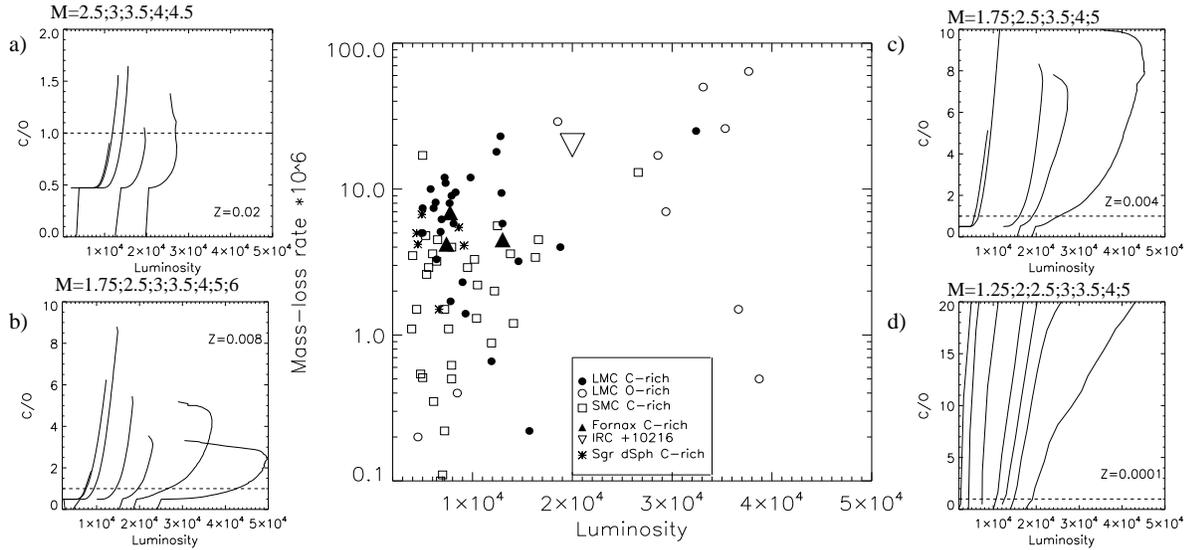}
\caption{\label{mass_loss_co}Mass-loss rates as a function of luminosity for 
different Local Group AGB stars.  Panels a,b,c and d show the evolution of the 
C/O ratio with luminosity using the models described in
 Karakas \& Lattanzio (2007.). Values of the different initial masses in 
solar units are on top of each panel.}
\end{figure*}

The mass-loss rates of stars observed with Spitzer have been estimated
by fitting spectra using a 1-d radiative transfer code, including
amorphous carbon dust and silicate carbide dust. The fitting process
and results are described in Groenewegen et al. (2007) for the MCs
stars, Matsuura et al. (2007) for the Fornax stars and Lagadec et
al. (2008) for Sgr dSph.

The results for the LMC, SMC, Sgr dSph and Fornax are shown in
Fig.\ref{mass_loss_co}.  We added the O-rich stars sample from van
Loon et al. (1999) and the well-studied Galactic carbon star
IRC+10216.  The figure shows that the mass-loss rates are highest for
the LMC stars, and tend to be lower for the SMC. This is under the
assumption of a constant gas to dust ratio. There is a tentative
indication that the gas-to-dust ratio is higher at the lowest
metallicities (Matsuura et al. 2007).

The mass-loss rates are around $10^{-5} \,\rm M_\odot \, yr^{-1}$
which is in the same range as the high mass-loss stars in the
Galaxy. A few Galactic stars may reach rates up to ten times higher,
but even in the Galaxy, such stars are very rare. Overall, there is no
evidence that peak (gas) mass-loss rates depend on Z for
carbon-rich stars.

\begin{table}
\caption[]{\label{cstars}Approximate numbers for life times and
magnitudes of carbon stars, derived from the current samples. $M_{\rm
bol}^{\rm init}$ indicates the magnitude where the large majority of
stars have become carbon-rich: i.e, the luminosity function becomes
flat. $M_{\rm bol(SW)}$ indicates the onset of the
superwind. Life times of the carbon-rich phase are taken from
Lagadec et al. (2007).  }
\begin{flushleft}
\begin{tabular}{llllllllllllllll}
\hline
        & [Fe/H]  &{$M_{\rm bol}^{\rm init}$} & life time & {$M_{\rm bol(SW)}$}\\
        &  & [mag] & [yr] & [mag] \\ 
\hline
 LMC &$-0.3$ & $-4.3$ & $3 \times 10^5 $ & $-4.7$ \\
 SMC &$-0.7$ & $-3.9$ & $6 \times 10^5 $ & $-4.5$ \\
 Sgr dSph &$-0.55$ & -- & --  & $-4.5:$ \\
\hline \\
\end{tabular}
\end{flushleft}
\end{table}

\subsection{Carbon star formation}

One unexpected result of the Spitzer surveys is the almost complete
dominance of carbon stars among the mass-losing stars in the
Magellanic Clouds. The original selection criteria did not separate
the two classes, so that this result shows a real effect. Part of the
explanation is that, at lower metallicity, more stars become carbon
stars.  Less dredge-up is required to overcome the original oxygen
abundance, and acquire the C/O$>1$ needed to form a carbon star. Based
on the luminosities, the progenitor masses of the mass-losing carbon
stars are $\sim$1.5--2.5 M$_{\odot}$ (Zijlstra et al. 2006).

The picture that emerges from this is that, at LMC metallicity, all
stars in the $\sim$1.5--2.5 M$_{\odot}$ range are C stars by the time
they develop substantial mass loss rates and they remain C stars until
their AGB evolution is terminated. The more luminous and massive, but
rare, OH/IR stars in the LMC ([e.g. Marshall et al. 2004) can survive
as O-rich objects well into the dust-enshrouded superwind phase,
possibly to the end of their AGB evolution.

Table \ref{cstars} lists the bolometric magnitude at which the onset
of the superwind occurs. This happens at a somewhat lower luminosity
($L\sim5\,10^3\,\rm L_\odot$) than is found in our Galaxy
($L\sim10^4\,\rm L_\odot$; $M_{\rm bol}\approx -5.5$). The difference
between the LMC and the Galaxy is that the former become carbon stars
at a point on the AGB where the latter are still oxygen-rich. We
interpret this as an indication that carbon-rich stars are more
efficient at mass loss, even where the oxygen-rich stars are more
metal-rich. The finding agrees with (but is less extreme than) the
prediction of Woitke (2006).

The few oxygen-rich stars in our MCs sample show mass-loss rates 
1-2 orders of magnitude less than the carbon stars (Sloan et al.,
2008). Thus, the mass-loss efficiency in oxygen-rich stars is more
affected by metallicity. This may not be unexpected: oxygen-rich dust
depends on metallicity-limited elements (Si, Al), while amorphous
carbon depends on self-produced carbon, and therefore the dust
abundance formation for oxygen-rich stars is much more strongly
dependent on metallicity. Taking the argument in reverse, the
implication would be that dust plays a role in driving the superwind
of oxygen-rich stars in the Galaxy.

To estimate the evolution of the carbon and oxygen abundances in the
circumstellar envelopes of the AGB stars we observed, we used the
stellar structure models described in Karakas \& Lattanzio (2007) and
made available for download. It calculates yields from detailed models
of low and intermediate-mass AGB stars with masses in the range 1--6
M$_{\odot}$. Models for four different metallicities are available:
$Z=0.02$ (solar), $Z=0.008$ (LMC), $Z=0.004$ (SMC) and $Z=0.0001$. We
have thus studied the evolution of the C/O abundance ratio with the
luminosity of the star for these four metallicities. At lower
metallicity, the initial oxygen abundance is lower and the stars
become carbon-rich earlier in their evolution. At solar Z,
stars with initial masses below 3 M$_{\odot}$ never become
carbon-rich, while in the LMC and SMC, this limit drops to 1.75 and
1.5 M$_{\odot}$ respectively.  For a star with a 2.5 M$_{\odot}$
initial mass, the C/O ratio becomes larger than 1 at $ L\sim 9.5
\,10^3\, \rm L_{\odot}$ in the LMC, $\sim 6.5\,10^3 \,\rm L_{\odot}$
in the SMC and $\sim 7.5\,10^3 \, \rm L_{\odot}$ for $Z=0.0001$. (In
the last case, the larger value is due to a metallicity effect in the
core-mass--luminosity relation). The evolution of the C/O ratio with
luminosity for the four different metallicities are shown in
Fig.\ref{mass_loss_co}

At low metallicity, third dredge-up is more efficient in making stars
carbon rich: the same amount of primary
carbon will have a larger effect on more metal-poor stars. This
process does not cease once the star has become carbon rich: further
dredge-up will continue to enhance the C/O ratio.  We may therefore
expect that metal-poor carbon stars have a higher C/O ratio than do
metal-rich carbon stars. Matsuura et al. (2005a) find evidence that
LMC stars have C/O ratios of $\sim 1.5$, versus $\sim 1.1$ for typical
Galactic carbon stars. Ratios in lower metallicity systems (e.g. SMC)
could be higher.

This immediately affects the potential amount of amorphous carbon
dust, which will depend on the amount of 'free' carbon. Instead of the
C/O ratio, a more relevant parameter is 
$$X_{\rm C} = \rm \frac{C-O}{O_\odot}$$ which measures the excess
carbon by number, scaled to the solar oxygen abundance. $X_{\rm C}$ rapidly
increases towards lower metallicity. The corresponding increase in the
C$_2$H$_2$ abundance is in fact seen, with the acetylene bands becoming much
stronger (larger equivalent width) in the sequence Galaxy--LMC--SMC (Lagadec
et al. 2007). In contrast, the absence of HCN bands suggests its abundance
becomes limited by nitrogen, which becomes enhanced only in the more massive
stars.

The dust formation efficiency may depend primarily on $X_{\rm
C}$. (The seeds required for the dust formation (e.g. TiC) may
introduce a dependency on metallicity.)  For mass loss driven by
radiation pressure on carbonaceous dust, we make the assumption that
{\it the superwind is triggered when a limiting value for $X_{\rm C}$
is reached}. For Galactic stars, this occurs for C/O$=1.1$ $(X_{\rm
C}=0.1)$ (Ohnaka et al., 2000). We predict that for the LMC the
superwind begins at C/O$=1.25$ and for the SMC, at C/O$=1.5$. The stronger 
acetylene bands at lower metallicity, and  the indicative
C/O ratios (Matsuura et al. 2005a) suggest similar but possibly slightly
higher C/O ratios.

\begin{figure}
\includegraphics[width=9cm]{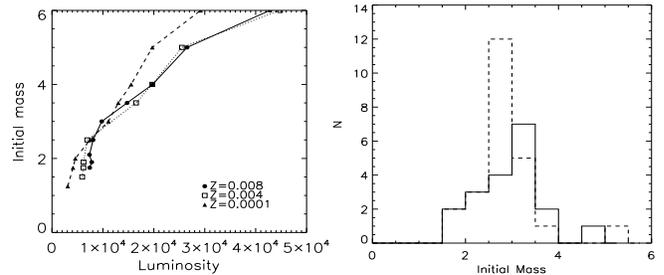}
\caption{\label{mass_lumi}Left:Initial mass of the stars as a function of the
  luminosity at which their C/O ratio becomes larger than 1. Right:
  Distribution of the initial mass of the AGB stars in our LMC (black line)and
  SMC sample (dashed line) as derived from the relation son the left. Values are in solar units.}
\end{figure}

\section{Discussion}
\begin{table*}
\begin{center}
\caption[]{\label{tab_lumi_trigger} The luminosities for which C/O reaches 1,
or X$_C$ 0.1 for different metallicities and initial masses, derived
from the stellar evolution models.  $\emptyset$ indicates that no
model was available and $-$
indicates that the considered star never become
carbon-rich. Luminosity are in unit of 10$^4$L$_{\odot}$.}
\begin{flushleft}
\begin{tabular}{lcccccccccccccccc}
\hline
  Mass (M$_{\odot}$)     & \multicolumn{2}{c}{Z=0.02}  &\multicolumn{2}{c}{Z=0.008}& \multicolumn{2}{c}{Z=0.004 }& \multicolumn{2}{c}{Z=0.0001}\\
        & L\,(C/0=1) & L\,(X$_C$=0.1)& L(\,C/0=1)& L\,(X$_C$=0.1) & 
L\,(C/0=1)& L\,(X$_C$=0.1)& L\,(C/0=1)& L\,(X$_C$=0.1)\\ 
\hline
1.0 &-&-      &-&-       &-&-       &-&-\\
1.25 &-&-     &-&-       &-&-       &0.31&0.31\\
1.5 &-&-      &-&-       &0.59&0.65 & $\emptyset$&$\emptyset$\\
1.75 &-&-     &0.74&0.79 &0.62&0.62 &0.41&0.41\\
1.9 &-&-      &0.78&0.78 &0.62&0.62 & $\emptyset$&$\emptyset$\\
2.0 &-&-      & $\emptyset$& $\emptyset$       &$\emptyset$ &   $\emptyset$     &0.46&0.46\\
2.1 &$\emptyset$ & $\emptyset$      &0.74&0.79 & $\emptyset$&$\emptyset$        &$\emptyset$ &$\emptyset$\\
2.5 &-&-      &0.81&0.81 &0.69&0.69 &0.75 &0.75\\
3.0 &1.21&1.24&0.98&0.98 & $\emptyset$&   $\emptyset$     &1.11&1.11\\
3.5 &1.45&1.47&1.48&1.48 &1.65&1.65 &1.30&1.30\\
4.0 &1.94&1.94&1.97&2.02 &1.97&1.97 &1.55&1.55\\
5.0 &2.69&2.69&2.65&2.84 &2.55&2.65 &1.97&1.97\\
6.0 &-&-      &4.28&4.40 &4.48&4.77 &2.92&3.07\\

\hline \\
\end{tabular}
\end{flushleft}
\end{center}
\end{table*}

For oxygen-rich stars, Galactic stars show a superwind for
luminosities of $\sim 10^4\,\rm L_\odot$.  The mass-loss rate of a
dust-driven wind scales as $L^{3/4}$ (van Loon, 2007) and is proportional to the dust-to-gas
mass ratio $\Psi$. As silicate
dust depends on Si and O, which are not enriched in AGB stars, we
assume that $\Psi$ scales as $Z$.  We now make the assumption that the superwind
starts when this relation predict a critical mass-loss rate.  Scaling from the
Galaxy, we then expect that an oxygen-rich dust-driven superwind (OSW)(larger
than $10^{-7}M_{\odot}$yr$^{-1}$) is
triggered for
$$ L_{\rm OSW} = 10^4 \left(\frac{Z}{Z_\odot}\right)^{-4/3}\,\rm
L_\odot$$ For the LMC and SMC this corresponds to $3.4 \times 10^4$
and $8.5 \times 10^4\,\rm L_\odot$, with an uncertainty of a factor of $\sim$2
due to the estimation of the value of $ 10^4\,\rm L_\odot$
and the metallicity dispersion within the galaxies . Fig. \ref{mass_loss_co} shows that
the least luminous OSW star in the LMC has $L \sim 1.8 \,10^4\,\rm
L_\odot$.  At lower luminosities, oxygen-rich stars have much lower
mass-loss rates than do carbon rich stars.

For luminosities below this limit, a superwind is seen only for
carbon-rich stars (Fig.\ref{mass_loss_co}).  There is no clear
evidence of increasing mass-loss rates with increasing luminosity.
The carbon-rich superwind (CSW) in low Z AGB stars occurs
even at luminosities below 1$\times$10$^4$L$_{\odot}$, showing the
efficacy of carbon dust in driving mass loss.

We thus propose that the superwind is triggered when either

i) the luminosity of an AGB star becomes larger than a luminosity
limit ($L_{\rm limit}\sim 10^4 (Z/Z_\odot)^{-4/3}\,\rm L_\odot$), regardless of
chemistry, over which the radiation pressure becomes dynamically
important. This is the dominant mechanism at solar metallicity

ii) or  the amount of free carbon reaches $X_{\rm C} \sim$0.1, when a CSW starts.
 This is the dominant mechanism at subsolar metallicity, for stellar masses
 above the limit for third dredge up.

Pulsation-driven winds are not included here: we expect these may
become important for low-metallicity, low-mass stars which do not
experience third dredge-up.

Fig.\ref{mass_lumi} shows the relation between the initial mass of the
star and the luminosity at which X$_C \sim$0.1 according to the
stellar models.  If the superwind is triggered when X$_C \sim$0.1,
then we can estimate the initial mass of the observed stars using this
relation.  The histograms of the observed stars in the LMC and SMC are
shown in Fig.\ref{mass_lumi}. These histograms
show that most of the observed stars have an initial mass between 1.5
and 3.5 M$_{\odot}$.  If
X$_C$ is the main factor to trigger the superwind, then we can also
predict the luminosities at which the superwind will be triggered for
different masses and metallicities. These values are shown in Table
\ref{tab_lumi_trigger}.
 
\section{Conclusions}
In this letter, we made use of a Spitzer spectroscopic survey of
mass-losing AGB star in Local Group low metallicity galaxies to study
the superwind from the stars. We have shown that the luminosity at
which the superwind is observed is decreasing with lower metallicity for
carbon-rich stars. Oxygen-rich stars in the Galaxy and in lower
metallicity galaxies have similar mass-loss rates for similar
luminosities.  To explain this dichotomy, we propose that the main
agent to trigger the superwind in low metallicity carbon stars is the
abundance of carbon. When the C/O ratio becomes larger than unity,
carbon-rich dust forms and the radiation pressure on this dust becomes
more efficient.  We used stellar evolution models to check the
consistency of our explanations and made some prediction of the
luminosities at which the superwind should be triggered for different
metallicities and stars initial masses.

\section*{Acknowledgments}
EL acknowledges support from a STFC rolling grant.
We thanks Amanda Karakas for making her results available and for her help as
well as 
the referee, Jan-Martin Winters, for his comments who helped
improving the quality of the letter.

\label{lastpage}

\end{document}